\documentclass[prd,tightenlines,floatfix,showpacs,preprintnumbers,nofootinbib,eqsecnum,superscriptaddress,preprint]{revtex4-1}

\usepackage{epsfig}
 
\def\d{{\rm d}}

\def\lr{\left( }
\def\rr{\right) }
\def\le{\left[ }
\def\re{\right] }

\def\beq{\begin{equation}}
\def\eeq{\end{equation}}
\def\bea{\begin{eqnarray}}
\def\eea{\end{eqnarray}}
\def\ms {\overline{\rm MS}}
\def\dis{{\rm DIS}_{\gamma}}

\begin{document}

\preprint{MS-TP-18-06}

\title{Nuclear parton density functions from dijet photoproduction \\ at the EIC}

\author{M.\ Klasen}
\email{michael.klasen@uni-muenster.de}
\affiliation{Institut f\"ur Theoretische Physik, Westf\"alische
 Wilhelms-Universit\"at M\"unster, Wilhelm-Klemm-Stra{\ss}e 9,
 D-48149 M\"unster, Germany}

\author{K.\ Kova\v{r}\'ik}
\email{karol.kovarik@uni-muenster.de}
\affiliation{Institut f\"ur Theoretische Physik, Westf\"alische
 Wilhelms-Universit\"at M\"unster, Wilhelm-Klemm-Stra{\ss}e 9,
 D-48149 M\"unster, Germany}

\date{\today}


\begin{abstract}
We study the potential of dijet photoproduction measurements at a future
electron-ion collider (EIC) to better constrain our present knowledge of the
nuclear parton distribution functions. Based on theoretical calculations
at next-to-leading order and approximate next-to-next-to-leading order of
perturbative
QCD, we establish the kinematic reaches for three different EIC designs, the
size of the parton density function modifications for four different
light and heavy nuclei from He-4 over C-12 and Fe-56 to Pb-208 with respect
to the free proton, and the improvement of EIC measurements with respect to
current determinations from deep-inelastic scattering and Drell-Yan data alone
and when also considering data from existing hadron colliders.
\end{abstract}


\pacs{
12.38.Bx, 
13.60.Hb, 
13.87.Ce, 
24.85.+p 
}

\maketitle


\section{Introduction}
\label{sec:1}

Our present knowledge about the structure of hadrons at high energies is
mostly encoded in parton density functions (PDFs). Since only the evolution
of these quantities with the energy scale $Q$ can be calculated in
perturbative QCD, but not their dependence on the longitudinal parton momentum
fraction $x$,
they are generally fitted to experimental data using factorization theorems
and calculations of the Wilson coefficients at next-to-leading order (NLO)
and beyond \cite{Collins:1989gx}. The classical process for the extraction
of PDFs is inclusive deep-inelastic scattering (DIS). Combined measurements of
this process by the H1 and ZEUS experiments at DESY HERA have led to precise
determinations of the proton PDFs \cite{Abramowicz:2015mha}.

Since the gluon density enters inclusive DIS only at NLO, other processes
with leading order (LO) gluon contributions such as inclusive jet or dijet
production in DIS \cite{Klasen:1997jm} and photoproduction \cite{Klasen:1995ab}
are also important. Today, data from DESY HERA and earlier experiments are
complemented by CERN LHC data on dijet, heavy-quark and electroweak
boson production \cite{Dulat:2015mca}. Understanding the structure of the
proton ($p$) is not only an interesting research topic in its own right, but
is also important to reliably estimate the production cross sections for new
particles and their backgrounds \cite{Butterworth:2015oua}.

For nuclei ($A$), experimental information on their PDFs came until very
recently almost exclusively from neutral and charged current fixed-target DIS
as well as Drell-Yan (DY) experiments, which limited the kinematic reach to
Bjorken $x$-values above about $10^{-2}$ and $Q^2$ values below $10^2$
GeV$^2$. The uncertainties of global nuclear PDF (nPDF) fits were therefore
considerably larger than they were for protons \cite{Hirai:2007sx}. In
particular, very little is known on the gluon PDF in nuclei, which is,
however, important to understand nuclear shadowing \cite{Armesto:2006ph}, its
possible relation to diffraction \cite{Frankfurt:2003zd}, saturation
\cite{GolecBiernat:1998js} and the initial condition for the creation of the
quark-gluon plasma in heavy-ion collisions \cite{Gelis:2010nm}. The situation
could be somewhat improved by including pion production data from BNL RHIC
\cite{Kovarik:2015cma}, albeit at the cost of introducing a fragmentation
function uncertainty, and recently also with first electroweak boson
\cite{Brandt:2013hoa} and in particular dijet \cite{Armesto:2015lrg} data from
pPb collisions at the CERN LHC \cite{Eskola:2016oht}. In addition, a recent
reweighting study has shown that also forward heavy-quark and quarkonium
production data from the CERN LHC have the potential to better constrain future
analyses \cite{Kusina:2017gkz}.

A future electron-ion collider (EIC) combining a new electron beam with the
existing high-energy Relativistic Heavy-Ion Collider RHIC (eRHIC)
\cite{mueller} or a new ion beam with the existing high-luminosity Continuous
Electron Beam Accelerator Facility (CEBAF) at Jefferson Lab at medium energy
(MEIC) \cite{yoshida} now offers the opportunity for measurements of nPDFs
that can reach and surpass the precision known from DESY HERA. The impact of
inclusive DIS has already been studied in 2012 in a White Paper
\cite{Accardi:2012qut}, which was recently updated based on newer nPDFs. As
shown there, an improvement of up to an order of magnitude in precision can be
expected in inclusive DIS at low $x$ \cite{Aschenauer:2017oxs}. In a
recent publication, we studied the impact of inclusive jet production
measurements in DIS, i.e.\ of photons with large virtuality $Q^2$, at the EIC
and reached similar conclusions \cite{Klasen:2017kwb} based on our previous
theoretical calculations at NLO \cite{Klasen:1997jm} and approximate
next-to-next-to-leading order (aNNLO) \cite{Biekotter:2015nra}. Full NNLO
calculations of inclusive jet \cite{Abelof:2016pby} and dijet production
\cite{Currie:2016ytq} in DIS are now also available. They confirm the aNNLO
results even at surprisingly large distances from hadronic threshold and
show that the NNLO corrections are moderate in size, except at the kinematical
edges, and that their inclusion leads to a substantial reduction of the scale
variation uncertainty on the predictions. 

Here, we focus on the complementary region of almost real photons with
$Q^2\simeq0$. Then not only direct, but also resolved photons contribute
\cite{Klasen:2002xb}, so that jet photoproduction at the EIC also has the
potential to finally better constrain the PDFs in the photon
\cite{Chu:2017mnm}. We consider dijet instead of inclusive jet photoproduction,
so that the probed $x$ values in the heavy ion and the photon can be
reconstructed (at LO exactly, beyond LO approximately) from the final state.
For our numerical study, we use our established theoretical formalism of NLO
calculations \cite{Klasen:1995ab}, which we have recently updated to include
also aNNLO contributions \cite{Klasen:2013cba} based on a unified approach to
threshold resummation that allows to obtain these contributions via a
perturbative re-expansion \cite{Kidonakis:2003tx}.

We present the kinematic
reach of dijet photoproduction for three different currently discussed
configurations of the EIC, discuss the size of nuclear effects to be expected
for different light and heavy nuclei, estimate the improvement in sensitivity
on the nPDFs from the EIC with respect to current uncertainties, and establish
the size of the gluon contribution in the heavy nucleus and of direct vs.\
resolved contributions in the photon.

The remainder of the paper is organized as follows: In Sec.\ \ref{sec:2} we
review our theoretical formalism and in Sec.\ \ref{sec:3} the experimental
conditions that we consider. Sec.\ \ref{sec:4} contains our main numerical
results, and our conclusions and an outlook are given in Sec.\ \ref{sec:5}.


\section{Theoretical formalism}
\label{sec:2}

Thanks to the QCD factorization theorem \cite{Collins:1989gx}, the differential
dijet cross section in photoproduction can be expressed as
\beq
 \d\sigma=\sum_{a,b}\int\d y\,f_{\gamma/e}(y)\int\d x_\gamma \,
 f_{a/\gamma}(x_\gamma,\mu_\gamma)
 \int\d x_A \,f_{b/A}(x_A,\mu_A)\d\sigma_{ab}(\alpha_s,\mu_R,\mu_\gamma,\mu_A)\,.
\eeq
Here, 
\beq
 f_{\gamma/e}(y)=\frac{\alpha}{2\pi}\left[
 \frac{1+(1-y)^2}{y}\ln\frac{Q^2_{\max}(1-y)}{m_e^2 y^2}
 + 2 m_e^2 y\left(\frac{1}{Q^2_{\max}}-\frac{1-y}{m_e^2 y^2}\right)
 \right]
\eeq
is the improved Weizs\"acker-Williams flux for the bremsstrahlung of photons
with maximal virtuality $Q^2_{\max}$ and longitudinal momentum fraction $y$
from electrons with mass $m_e$ \cite{Frixione:1993yw}.

The photons can either
interact directly, so that at LO $f_{a/\gamma}(x_\gamma,\mu_\gamma)=
\delta(1-x_\gamma)$, or through their fluctuations into vector-meson like
quark-antiquark and gluon states described by the photon PDFs $f_{a/\gamma}
(x_\gamma,\mu_\gamma)$. Beyond LO, both contributions are related through the
factorization of collinear singularities. We use the GRV NLO parameterizations
of the photon PDFs \cite{Gluck:1991jc} obtained in the perturbatively stable
DIS$_\gamma$ scheme \cite{Gluck:1991ee}. These PDFs can be transformed to the
$\ms$ factorization scheme via
\beq
 f^{\ms}_{q/\gamma}(x_\gamma,\mu_\gamma) = f^{\dis}_{q/\gamma}(x_\gamma,\mu_\gamma)-{\alpha\over2\pi}\,e_q^2\, C_{\gamma}(x_\gamma)\,,
\eeq
i.e.\ through the absorption of the pointlike Wilson coefficient in the photon
structure function
\beq
C_{\gamma}(x)=3\, \left[ \left( x^2+(1-x)^2\right)
   \,\ln \, \frac{1-x}{x} + 8x(1-x)-1 \right]
\eeq
into the PDFs of quarks with fractional charge $e_q$ in the photon.
Subsequently, other NLO parameterizations of the photon PDFs in the $\dis$
\cite{Cornet:2004nb} and $\ms$ scheme \cite{Aurenche:2005da} have been
proposed. In the absence of experimental constraints, their spread must
be considered a contribution to the theoretical uncertainty that a future
EIC might also help to
reduce \cite{Chu:2017mnm}. As we will see, the constraints on nuclear and
photon PDFs come from complementary kinematic regions.

For the nuclear PDFs $f_{b/A}(x_A,\mu_A)$, we consider the nCTEQ15 NLO fits
with their intrinsic nuclear mass dependence and 32 associated error PDFs as
our baseline, and we estimate the impact of the inclusive pion production data
from BNL RHIC with their nCTEQ15-np variants \cite{Kovarik:2015cma}. In
addition,
we will show results using the more recent EPPS16 NLO fits, which are based on
the factorized form
\beq
 f_{b/A}(x_A,\mu_A)=R_{b/A}(x_A,\mu_A)\,f_{b/p}(x_A,\mu_A),
\eeq
information on the nuclear modification factor $R_{b/Pb}$ from pPb collisions
at the CERN LHC and CT14 NLO free proton PDFs $f_{b/p}(x_A,\mu_A)$
\cite{Dulat:2015mca}.

The partonic cross sections $\d\sigma_{ab}(\alpha_s,\mu_R,\mu_\gamma,\mu_A)$
are well known at NLO \cite{Klasen:1995ab}. We have recently included
approximate NNLO (aNNLO) corrections \cite{Klasen:2013cba} based on a unified
approach to NNLO soft and virtual corrections from a re-expansion of all-order
resummation \cite{Kidonakis:2003tx}. These corrections dominate close to
partonic threshold
\beq
 z={(p_1+p_2)^2\over(p_a+p_b)^2}~\to~1,
\eeq
i.e.\ when the invariant mass of the dijet pair with four-momenta $p_{1,2}$
approaches the one of the incoming partons with four-momenta $p_{a,b}$.
For brevity, we present here only the master formula at NLO
\beq
 \d\sigma_{ab}=\d\sigma_{ab}^B{\alpha_s(\mu_R)\over\pi}
 \le c_3D_1(z)+c_2D_0(z)+c_1\delta(1-z)\re
 +{\alpha_s^{d_{\alpha_s}+1}(\mu_R)\over\pi}\le A^cD_0(z)+T^c_1\delta(1-z)\re
\eeq
where the $+$-distributions
\beq
 D_l(z)=\le{\ln^l(1-z)\over 1-z}\re_+
\eeq
denote leading, next-to-leading logarithms etc.\ and $d_{\alpha_s}=0,1,2,...$
the power in the strong coupling constant $\alpha_s$ of the underlying Born
cross section $\d\sigma_{ab}^B$.
For a simple color flow, the second part of the equation is absent. The
master formula at NNLO and further details can be found in Ref.\
\cite{Kidonakis:2003tx}. For pair-invariant mass kinematics and in the $\ms$
scheme, the coefficients for a simple color flow read
\bea
 c_3&=&C_F-N_C,\nonumber\\
 c_2 &=& C_F\le-\ln\lr{\mu_A^2\over s}\rr-{3\over4}+2\ln\lr{-u\over s}\rr\re
     + N_C\ln\lr{t\over u}\rr-{\beta_0\over4},\nonumber\\
 c_1^\mu&=&-{3C_F\over4}\ln\left(\frac{\mu_A^2}{s}\right)
           +\frac{\beta_0}{4}\ln\left(\frac{\mu_R^2}{s}\right)
\eea
with $C_F=4/3$, $N_C=3$, $\beta_0=(11N_C-2n_f)/3$, $n_f$ quark flavors and the
usual Mandelstam variables $s$, $t$ and $u$ for the QCD Compton process
$\gamma q\to qg$, and
\bea
 c_3&=&2(N_C-C_F),\nonumber\\
 c_2 &=& -{3C_F\over2}
      +  N_C\le-\ln\lr{\mu_A^2\over s}\rr+\ln\lr{tu\over s^2}\rr\re,\hspace*{28.5mm}\nonumber\\
 c_1^\mu &=& -{\beta_0\over4}\ln\lr{\mu_A^2\over s}\rr
             +{\beta_0\over4}\ln\lr{\mu_R^2\over s}\rr
\eea
for photon-gluon fusion $\gamma g\to q\bar{q}$ \cite{Klasen:2013cba}. For
a complex color flow,
\bea
 c_3&=&2C_F,\nonumber\\
 c_2 &=& -C_F\ln\lr{\mu_\gamma^2\over s}\rr-C_F\ln\lr{\mu_A^2\over s}\rr-{11\over2}C_F,\nonumber\\
 c_1^\mu &=& -C_F\le\ln\lr{-t\over s}\rr+{3\over4}\re\ln\lr{\mu_\gamma^2\over s}\rr
 -C_F\le\ln\lr{-u\over s}\rr+{3\over4}\re\ln\lr{\mu_A^2\over s}\rr
 +{\beta_0\over2}\ln\lr{\mu_R^2\over s}\rr
\eea
for quark-(anti-)quark scattering $qq'\to qq'$ and $q\bar{q}'\to q\bar{q}'$ and
similarly for the other partonic processes \cite{Kidonakis:2003tx}. Note that
in the coefficients of the resolved processes also the photon factorization
scale $\mu_\gamma$ enters and that the coefficients $c_1^\mu$ contain only the
scale-dependent parts, whereas their finite parts must be taken from our full
NLO calculation \cite{Klasen:1995ab}.

The size of the aNNLO corrections has been shown not to exceed $+7$\% ($-7$\%)
at large jet transverse momentum $p_T$ and forward (backward) rapidity $\eta$
\cite{Klasen:2013cba}. More important is the reduction of the scale uncertainty
in particular at large $p_T$, which strengthens our confidence in the
perturbative calculation. Strictly speaking, the aNNLO formalism described
above applies to massless jets \cite{deFlorian:2013qia}, whereas experimentally
jets are defined with an algorithm and have non-vanishing mass. Work on
implementing the jet mass corrections is currently in progress. Their
impact is expected to be small, in particular when the jet radius $R=1$, so
that $\ln R$-terms vanish.


\section{Experimental conditions}
\label{sec:3}

Several variants of the EIC are currently under debate. The eRHIC version
proposed at BNL would add a new electron ring with energy $E_e=16\,...\,21$
GeV to the existing ion beam with energy $E_A=100$ GeV, so that a total
center-of-mass energy per nucleon of $\sqrt{s}=80\,...\,90$ GeV and an annual
integrated luminosity of about ${\cal L}=10$ fb$^{-1}$ could be reached.
At Jefferson Lab, the MEIC planning is built on the existing electron ring
with energy $E_e=12$ GeV and would add to it an ion beam of energy $E_A=40$
GeV, resulting in a lower center-of-mass energy of $\sqrt{s}=45$ GeV, but a
higher integrated luminosity of ${\cal L}=100$ fb$^{-1}$. We will therefore
consider all three collider scenarios.

The maximum virtuality $Q^2_{\max}$ and longitudinal momentum fraction $y$
of the photon can be determined either from the (anti-)tagged scattered
electron or from the hadronic final state with the Jacquet-Blondel method,
which has proven advantageous at very low values of $y$ at DESY HERA. Current
detector designs aim at $Q^2<$ 0.1 GeV$^2$ and 0.01 $\leq y\leq$ 0.95. The
electromagnetic
calorimeter would span the rapidity range $-4<\eta<4$ \cite{Accardi:2012qut}.
No specifications have so far been fixed for the hadronic calorimeter, so that
we assume the same coverage. In the following section we will, however,
see that a hadronic calorimeter of size $-1<\eta<3$ would be sufficient for
the jet measurements proposed here. We assume that the jets are reconstructed
with an anti-$k_T$ algorithm, a distance parameter $R=1$ in the $\eta-\phi$
plane, and a massless $p_T$ recombination scheme \cite{Cacciari:2008gp}.
Similarly to our study of inclusive jet production in DIS
\cite{Klasen:2017kwb}, we assume that jets can be measured down to transverse
momenta of $p_T\geq5$ (4.5) GeV, where the cuts on the leading (subleading)
jet must be sufficiently different to avoid sensitivity to soft gluon
radiation \cite{Klasen:1995xe}. We then set all scales to the average
transverse momentum $\mu_{R,\gamma,A}=\bar{p}_T=(p_{T,1}+p_{T,2})/2$.


\section{Numerical results}
\label{sec:4}

We now present our numerical results for dijet photoproduction at an EIC.
In particular, we compute single-differential cross sections in the average
momentum $\bar{p}_T$ and rapidity $\bar{\eta}=(\eta_1+\eta_2)/2$ of the two
jets as well as in the deduced initial parton momentum fractions
\bea
 x_{A}^{\rm obs} = {p_{T,1}\,e^{\eta_1}+p_{T,2}\,e^{\eta_2}\over 2E_A}
 &\quad{\rm and}\quad&
 x_{\gamma}^{\rm obs} = {p_{T,1}\,e^{-\eta_1}+p_{T,2}\,e^{-\eta_2}\over 2yE_e}
\eea
in the nucleus $A$ and the photon $\gamma$.

\subsection{Dijet photoproduction at different EICs}

\begin{figure}
 \epsfig{file=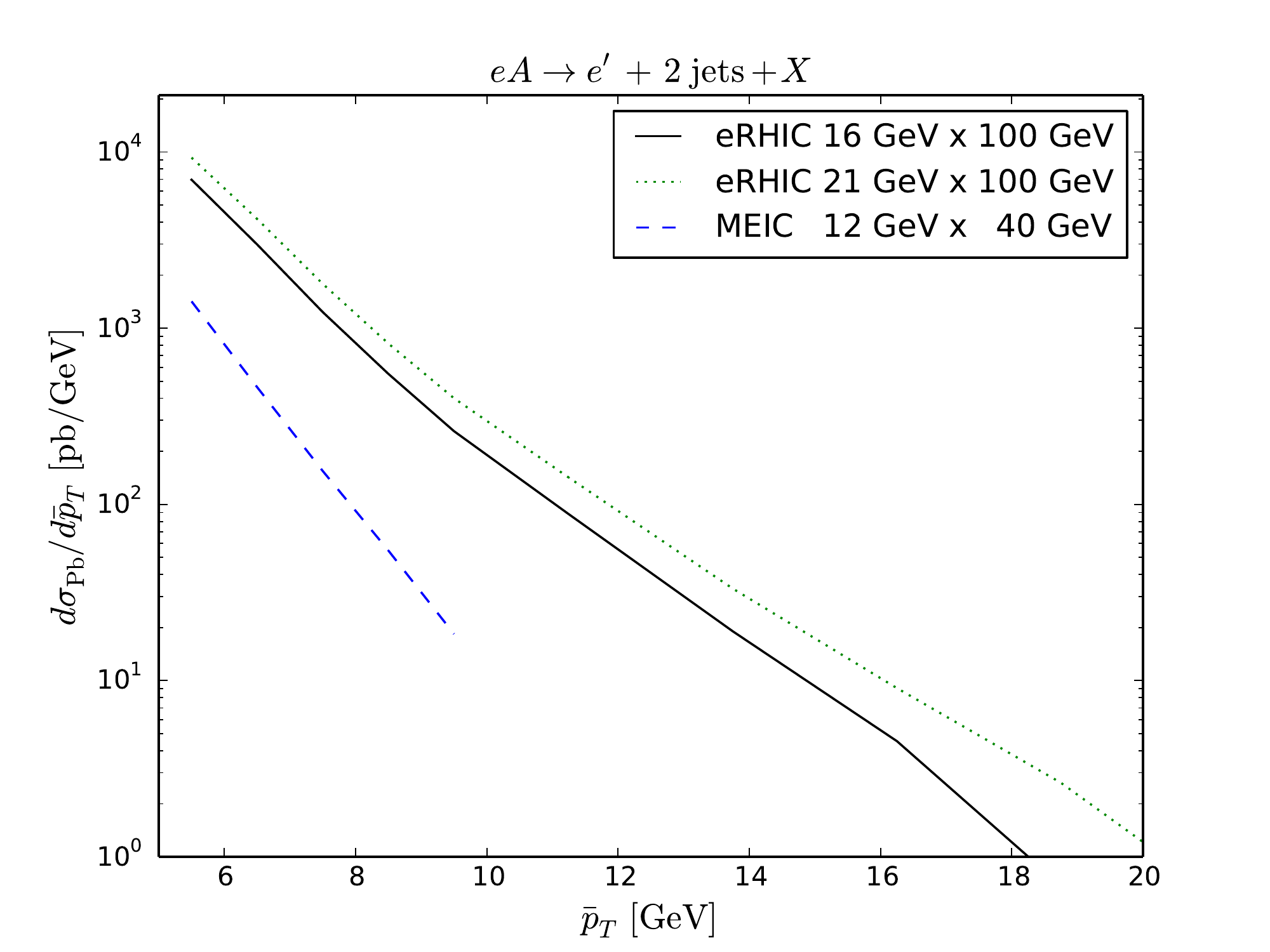,width=0.48\textwidth}
 \epsfig{file=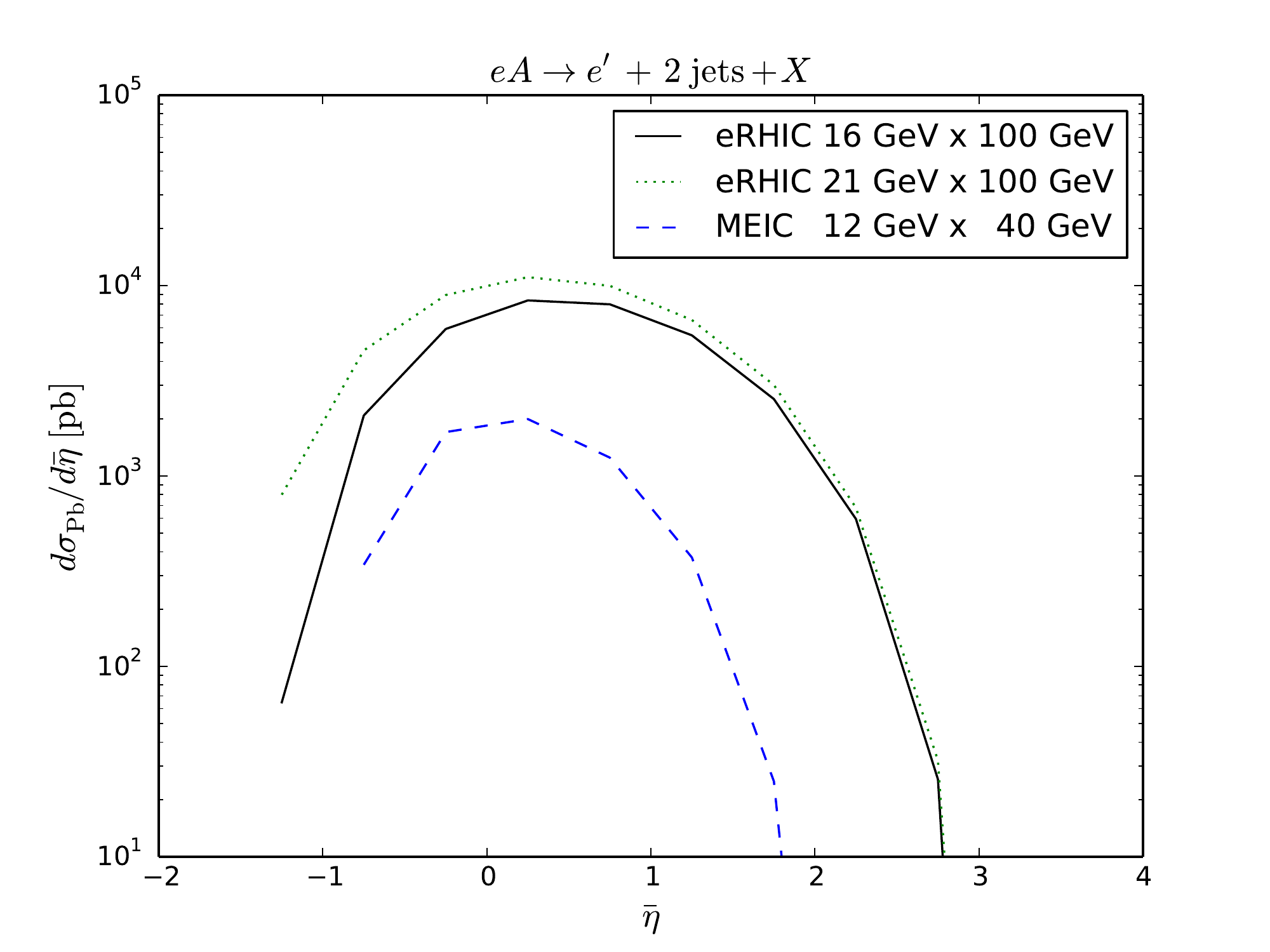,width=0.48\textwidth}
 \epsfig{file=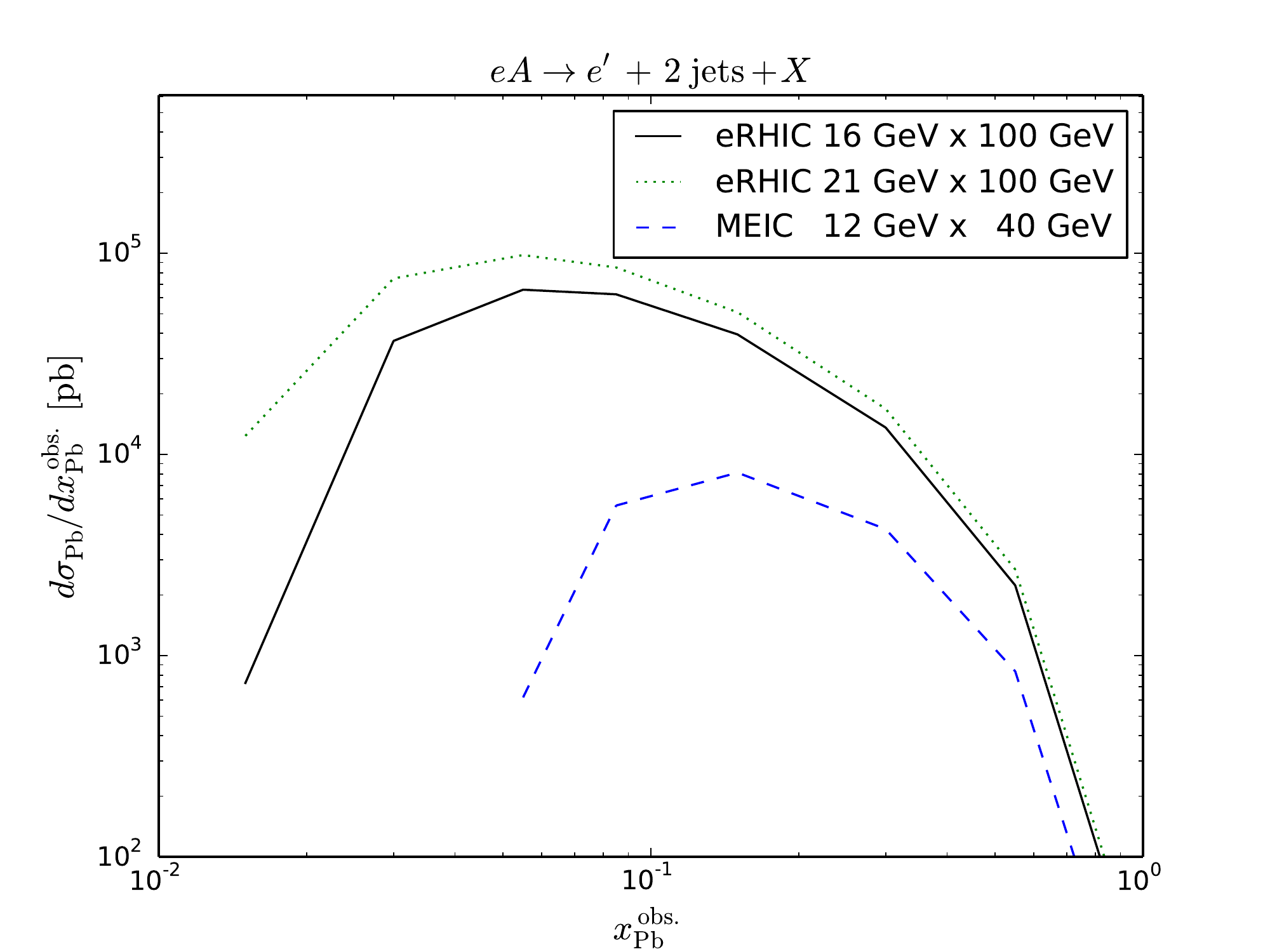,width=0.48\textwidth}
 \epsfig{file=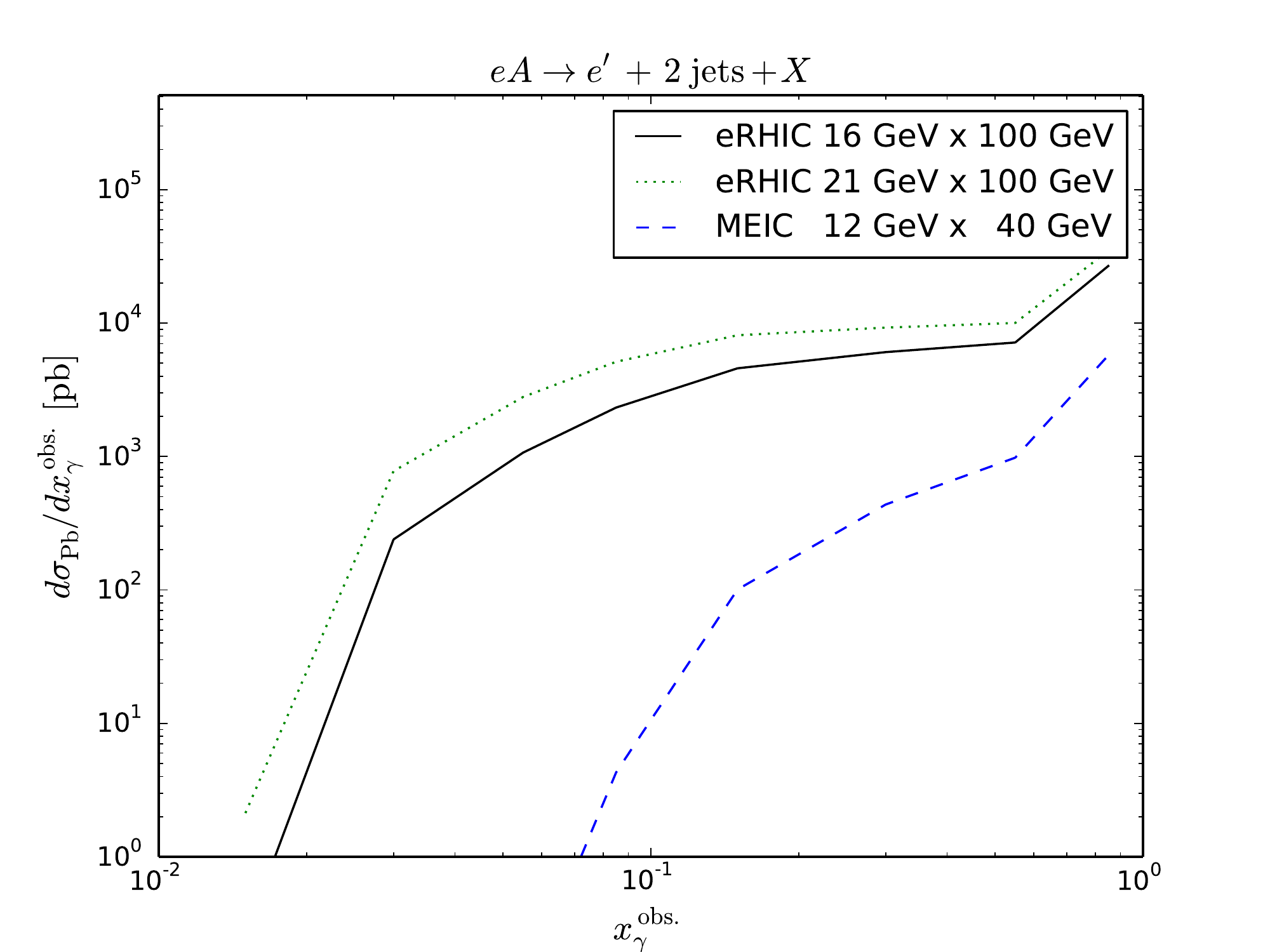,width=0.48\textwidth}
 \caption{\label{fig:1}Dijet photoproduction in electron-lead ion collisions
 at eRHIC and MEIC with electron beam energies of 12 to 21 GeV and ion beam
 energies per nucleon of 40 to 100 GeV. Shown are differential cross sections
 in the average jet transverse momentum (top left), average rapidity (top
 right), and observed parton momentum fractions in the probed nucleon (bottom
 left) and photon (bottom right).}
\end{figure}

Fig.\ \ref{fig:1} shows the kinematic reaches of the three EIC variants
discussed in the previous section. Average transverse momenta (top left) of up
to 10 and 20 GeV can be reached at MEIC and eRHIC, respectively, extending the
range in the probed scales by up to a factor of four from 100 to 400 GeV$^2$.
At the largest $\bar{p}_T$, between $10^3$ and $10$ events would be collected
annually with luminosities of 100 or 10 fb$^{-1}$. Compared to inclusive jet
production in DIS, where $p_T^2$ and $Q^2$ values of up to $10^3$ GeV$^2$ are
kinematically accessible \cite{Klasen:2017kwb}, the scales probed in dijet
photoproduction are therefore more limited.

Similar conclusions can be drawn for the $x$-values probed in the ion
(bottom left) and the photon (bottom right). At the MEIC, they cover the
region of the EMC effect \cite{Geesaman:1995yd} above $10^{-1}$ and
anti-shadowing \cite{Brodsky:2004qa} above a few $10^{-2}$, but do not reach
into the shadowing region below this value \cite{Armesto:2006ph,%
Frankfurt:2003zd}. The fact that the photon PDFs are only probed at larger
values of $x$ above a few $10^{-2}$, where they are dominated by the pointlike
(quark) contribution \cite{Klasen:2002xb} and gluon-initiated contributions
are small \cite{Chu:2017mnm}, is advantageous, as it reduces the photon PDF
uncertainty on the determination of the nuclear PDFs.

The two jets are produced with average rapidities (top right) between $-1$
and $2$ or $3$ at the MEIC or eRHIC. The ion beam is assumed to move
in the positive $z$ direction similarly to DESY HERA. This shows that a
hadronic calorimeter with this coverage would be sufficient to measure
dijet photoproduction.

\subsection{Dijet photoproduction on different nuclei}

\begin{figure}
 \epsfig{file=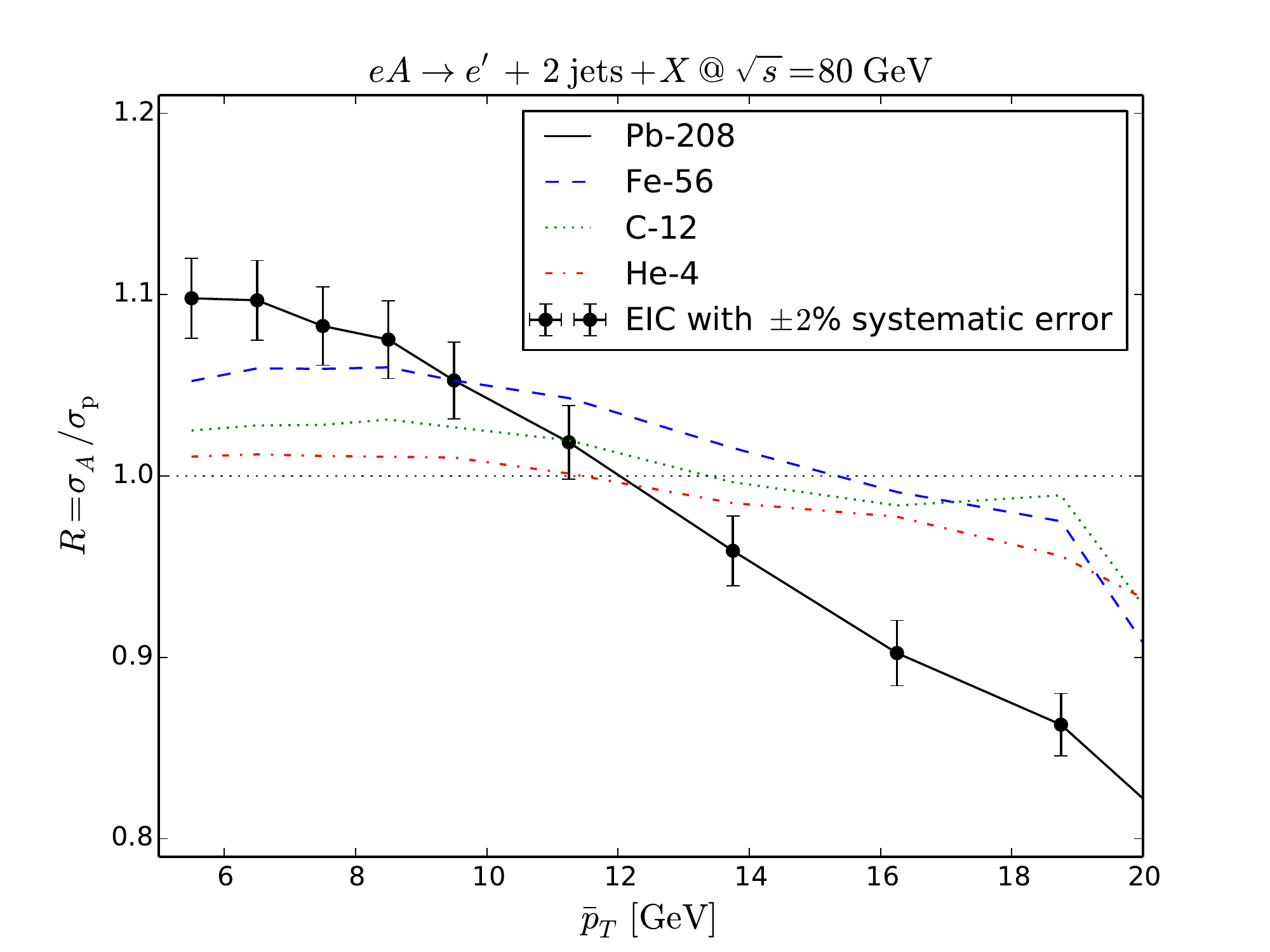,width=0.48\textwidth}
 \epsfig{file=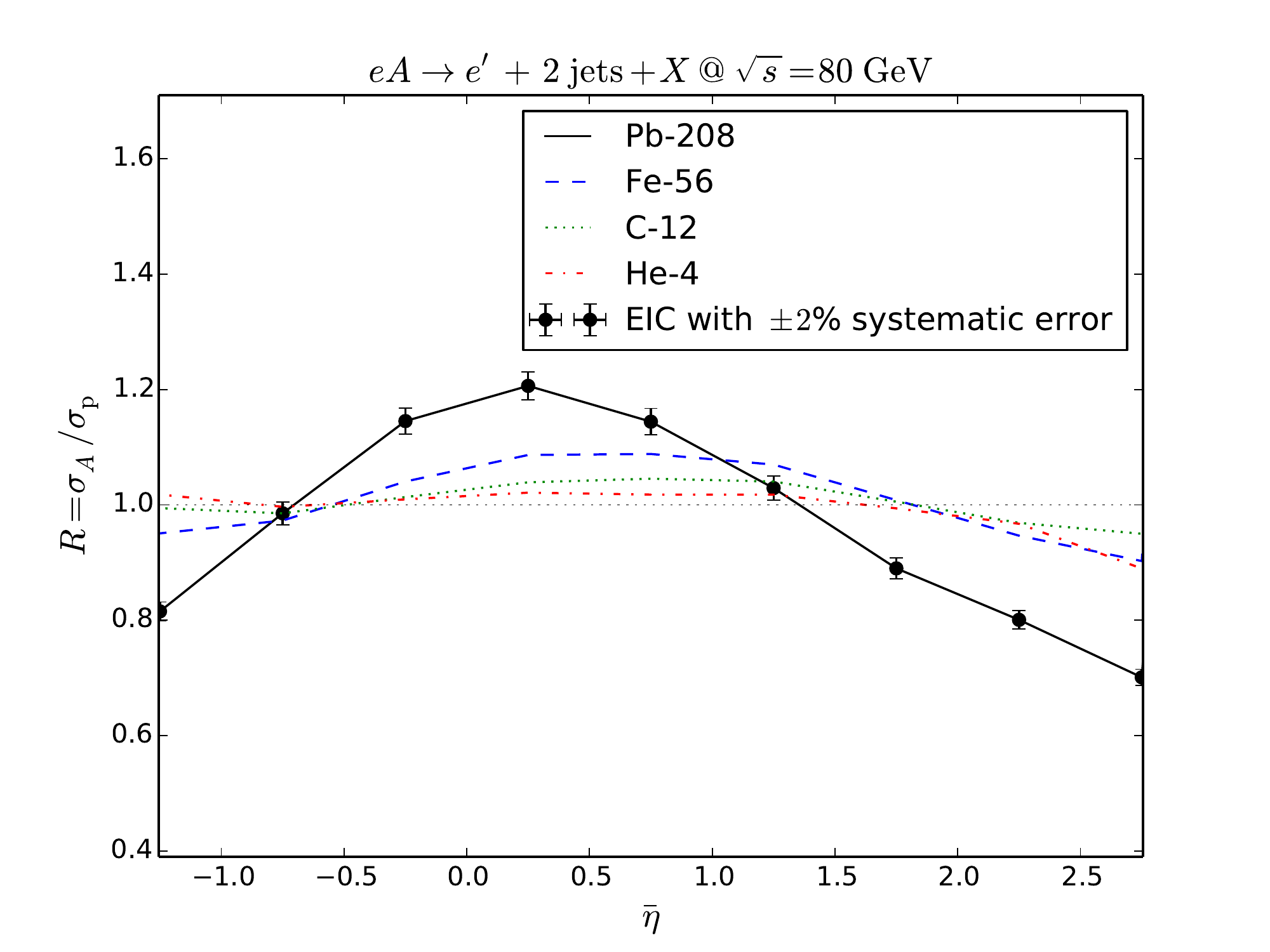,width=0.48\textwidth}
 \epsfig{file=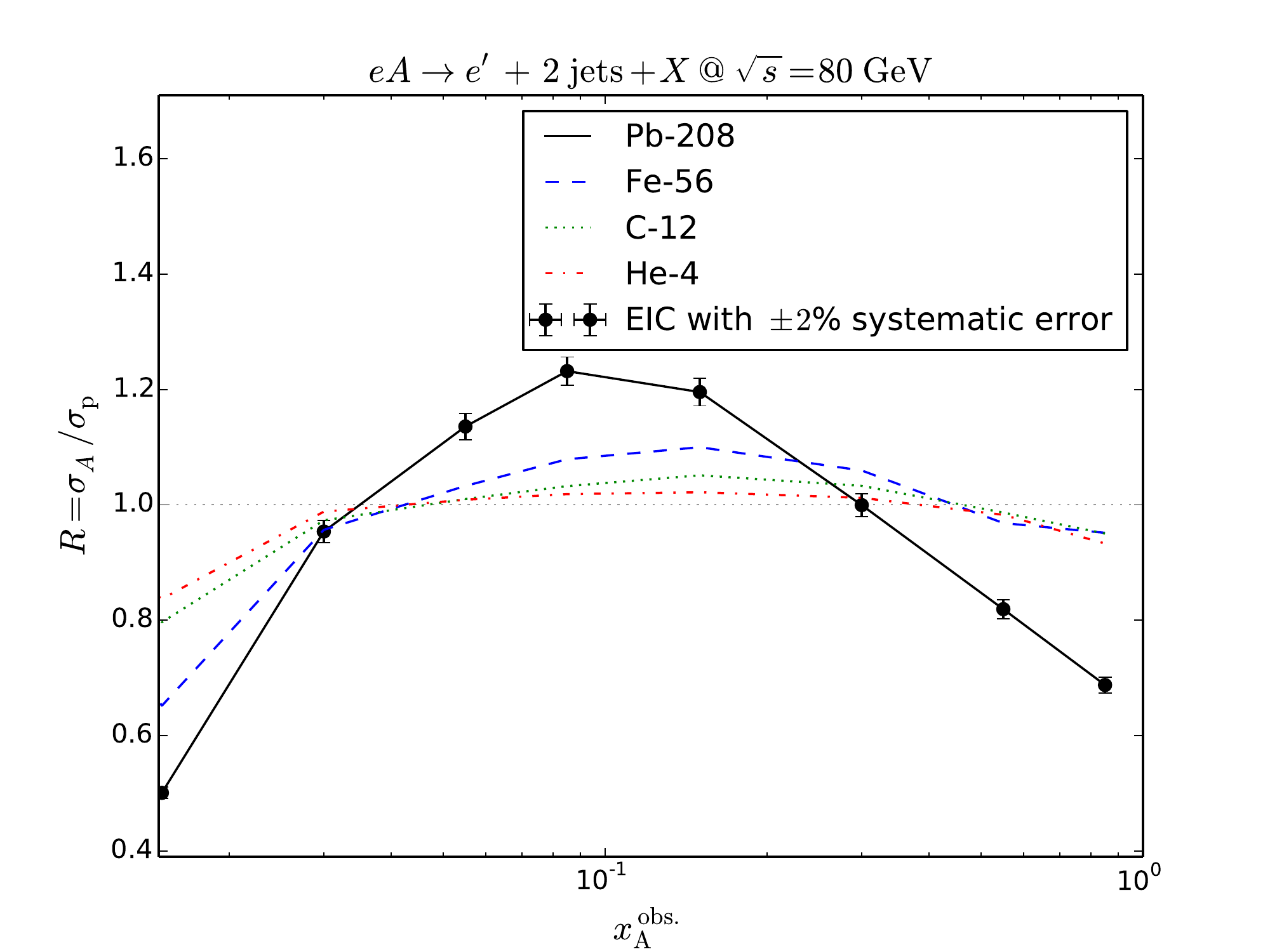,width=0.48\textwidth}
 \epsfig{file=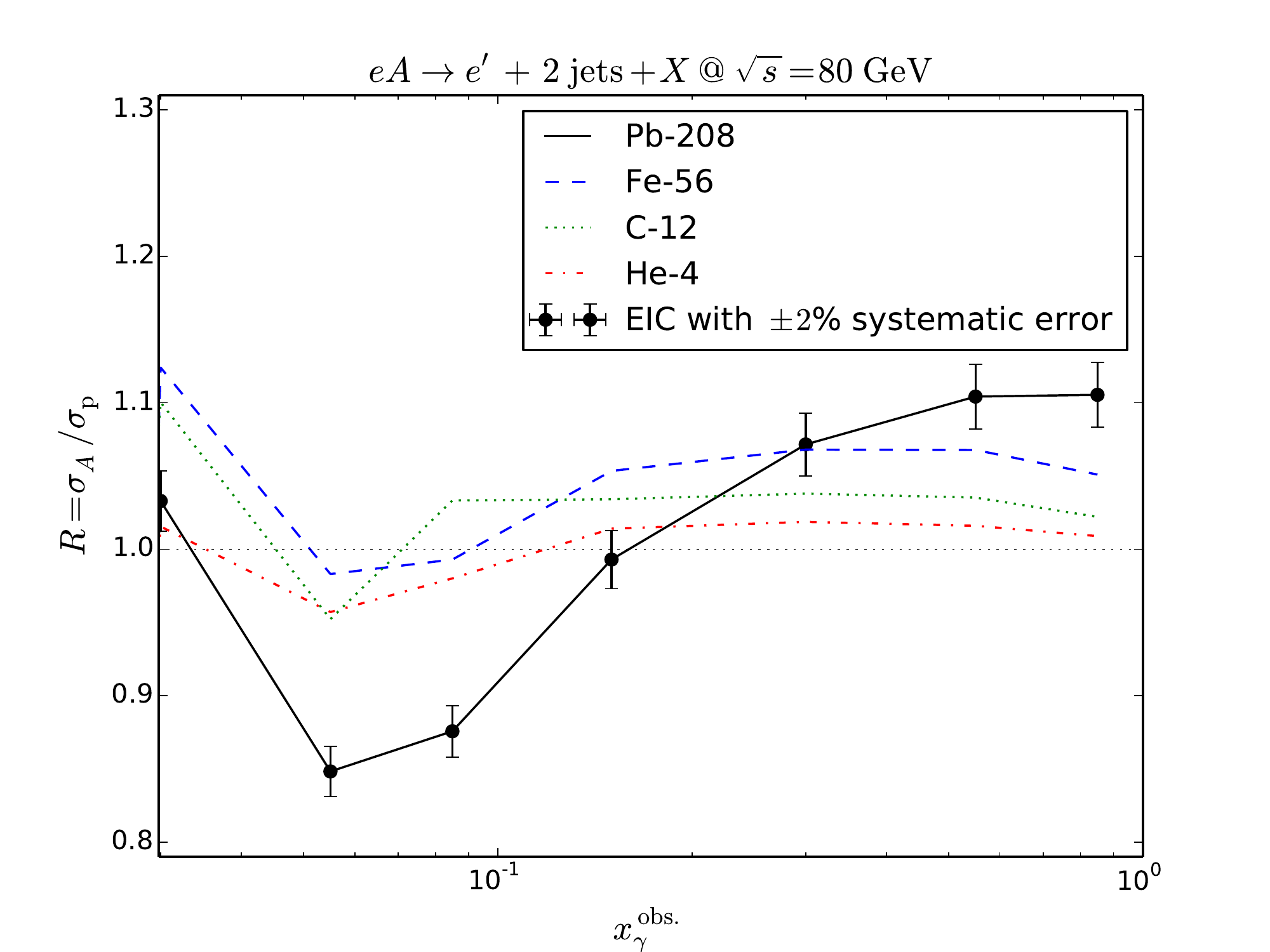,width=0.48\textwidth}
 \caption{\label{fig:2}Dijet photoproduction in electron-ion collisions with
 beam energies of 16 and 100 GeV, respectively, at eRHIC for different nuclei:
 Pb-208 (full black lines), Fe-56 (dashed blue lines), C-12 (dotted green
 lines), and He-4 (dot-dashed red lines). Shown are the ratios of electron-ion
 over electron-proton cross sections as a function of the average jet
 transverse momentum (top left), average rapidity (top right), and observed
 parton momentum fractions in the probed nucleon (bottom left) and photon
 (bottom right). Error bars indicate the expected experimental precision.}
\end{figure}

In this and the following subsections, we present ratios $R_A/R_p$ of
electron-ion over electron-proton cross sections as functions of the same
kinematic variables as above in order to study the sensitivity of the EIC
measurements on nuclear effects \cite{Arneodo:1992wf}. These ratios also
have the advantage of further reducing unphysical scale uncertainties.
We concentrate on
the eRHIC design with a center-of-mass energy per nucleon of $\sqrt{s}=80$
GeV. First, we study in Fig.\ \ref{fig:2} the size of these effets for
different light and heavy
nuclei from He-4 (dot-dashed red) over C-12 (dotted green) and Fe-56 (dashed
blue) to Pb-208 (full black line), based on the central nCTEQ15 fit. Generally
speaking, the difference to bare protons increases with the nuclear mass from
a few percent up to a factor of two. It changes sign from low to high
$\bar{p}_T$ (top left) and twice in the other distributions. The distribution
in $x_A^{\rm obs}$ (bottom left) clearly shows
the regions of the EMC effect, anti-shadowing and also shadowing at large,
intermediate and small $x$, which are correlated with the backward, central
and forward regions in $\bar{\eta}$ (top right). The distribution in
$x_\gamma^{\rm obs}$ (bottom right) shows that direct and pointlike photons,
which are well constrained, not only probe the shadowing, but also the
antishadowing region. On top of the prediction for lead ions,
we show simulated EIC data with a total systematic error of 2\% (black error
bars), which is expected to dominate over the statistcal error (cf.\ Fig.\
3.25 of Ref.\ \cite{Accardi:2012qut}).

\subsection{Sensitivity to nuclear parton density functions}

\begin{figure}
 \epsfig{file=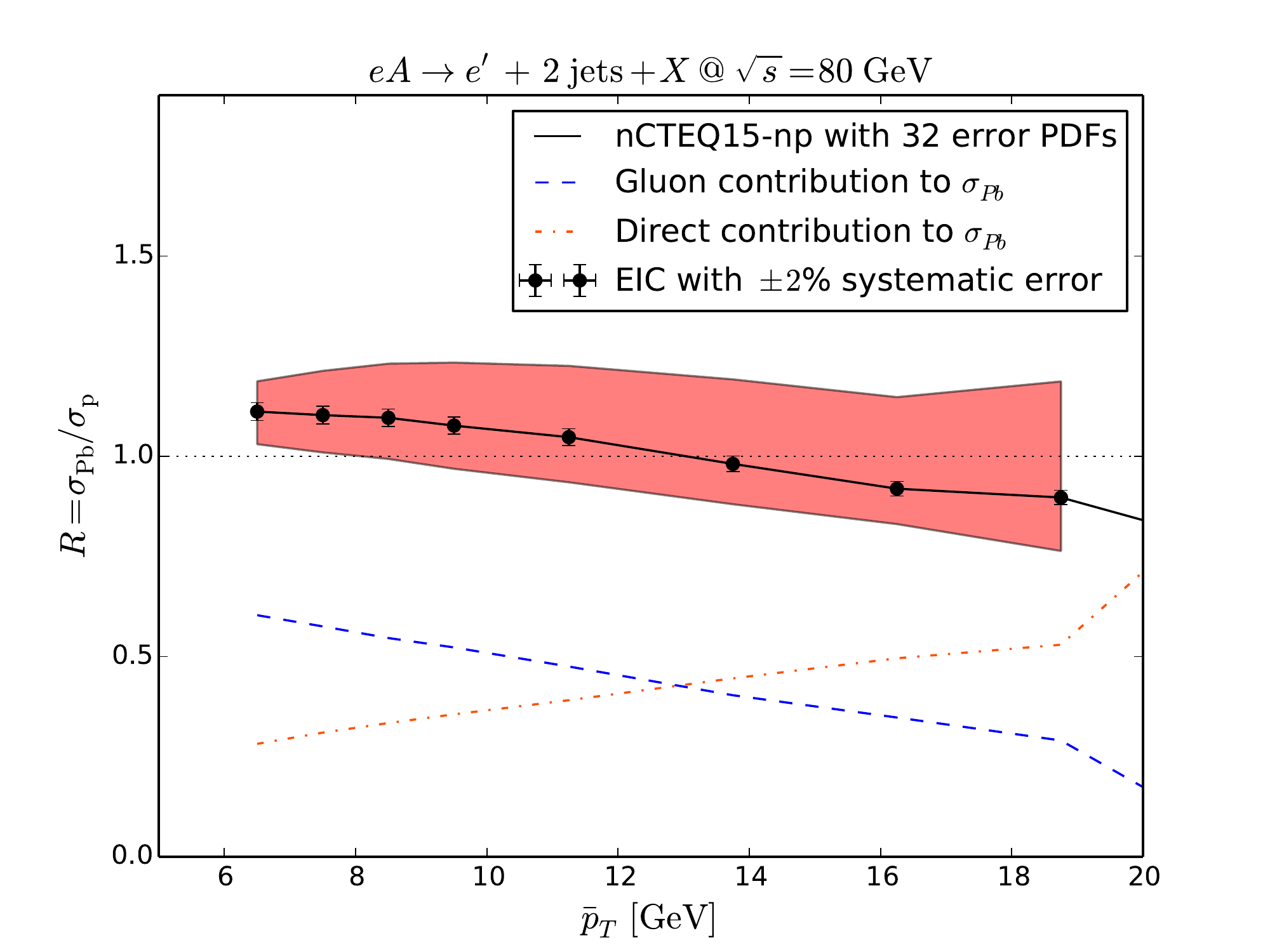,width=0.48\textwidth}
 \epsfig{file=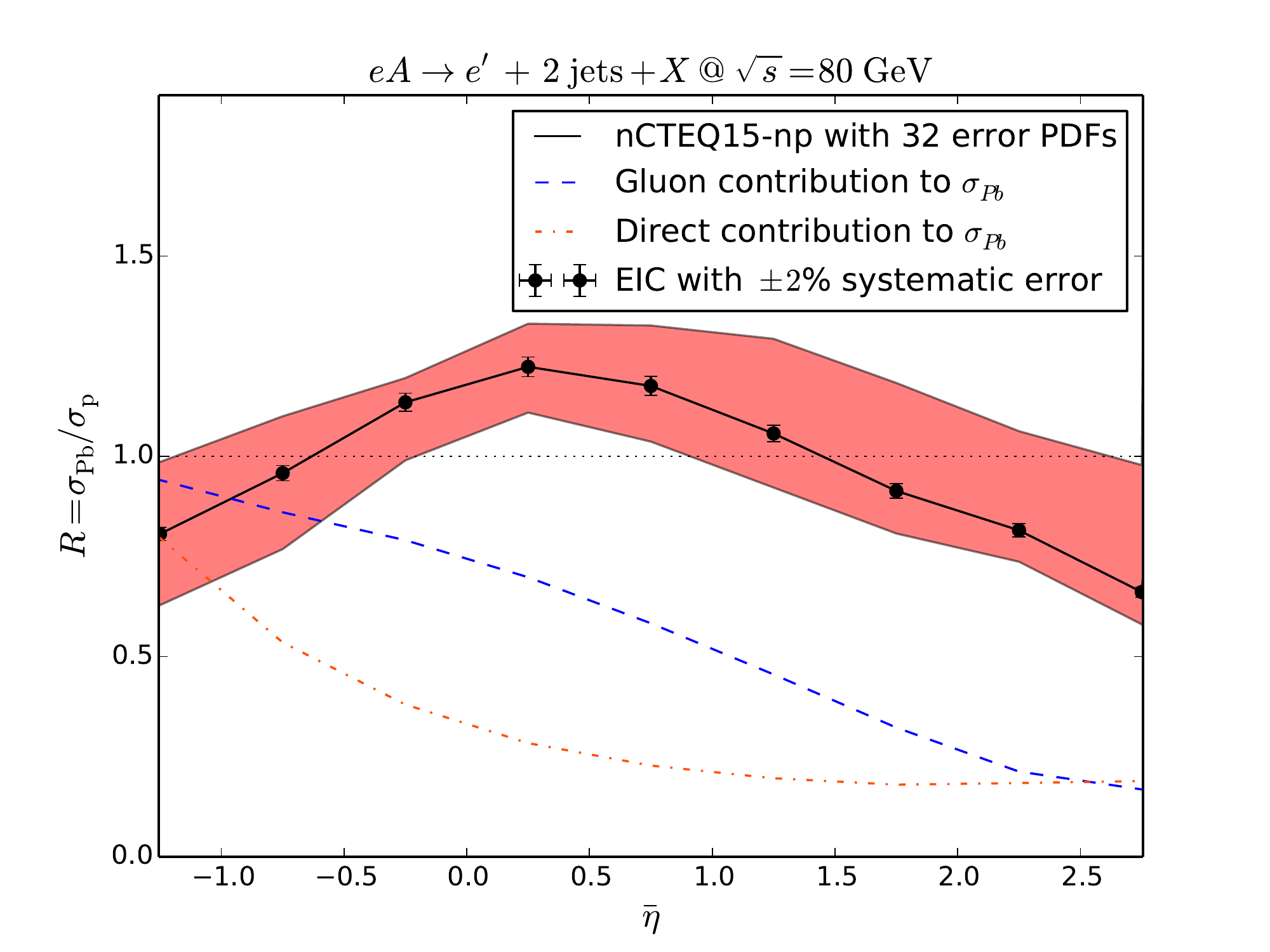,width=0.48\textwidth}
 \epsfig{file=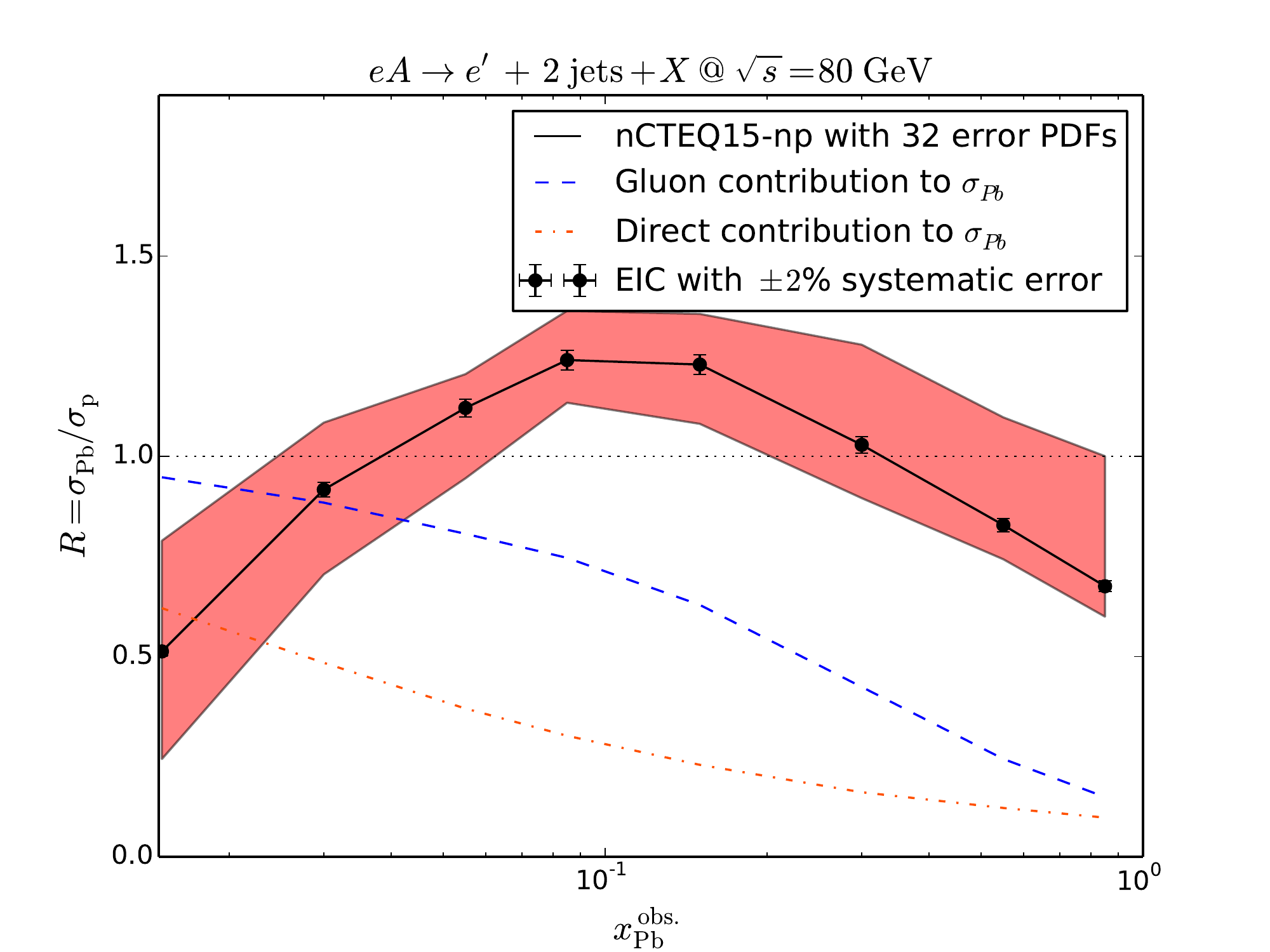,width=0.48\textwidth}
 \epsfig{file=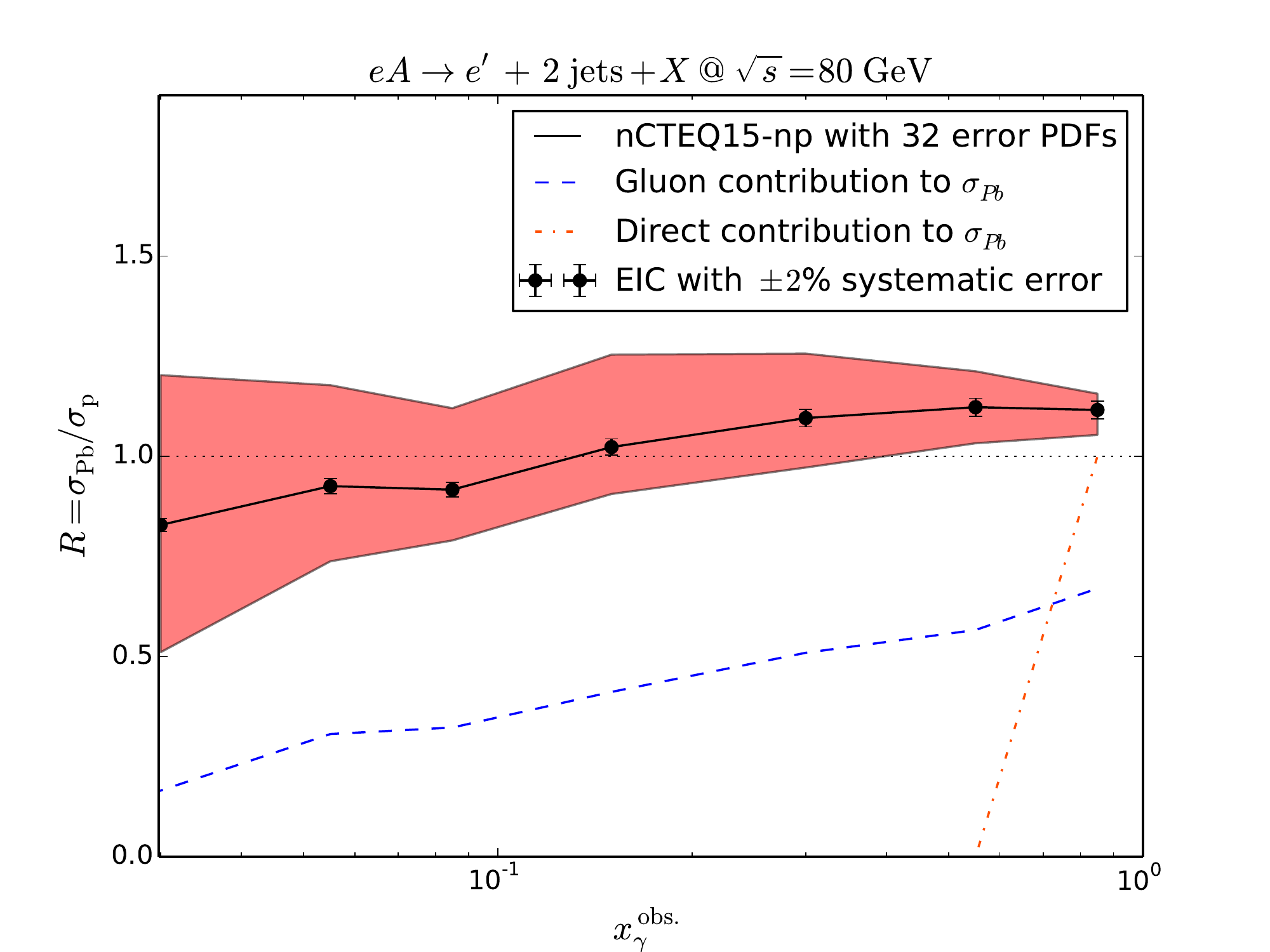,width=0.48\textwidth}
 \caption{\label{fig:3}Dijet photoproduction in electron-lead ion collisions
 with beam energies of 16 and 100 GeV, respectively, at eRHIC. Shown is the
 ratio of electron-lead ion over electron-proton cross sections (full black
 lines) including the current nuclear PDF uncertainty from the nCTEQ15 fit to
 DIS and DY data only (red-shaded bands) as well as the relative gluon
 contribution in the lead ion (dashed blue lines) and the direct photon
 contribution (dot-dashed orange) to the total cross section as a function of
 the average jet transverse momentum (top left), average rapidity (top right),
 and observed parton momentum fractions in the probed nucleon (bottom left)
 and photon (bottom right). Error bars indicate the expected experimental
 precision.}
\end{figure}

In Fig.\ \ref{fig:3} we focus on the predictions for Pb-208 and include the
current uncertainty of the nCTEQ15-np fit (red shaded bands), where no
constraints (and uncertainties) form pion production at BNL RHIC are included.
With the information from DIS and DY data alone, all four distributions are
consistent with unity within errors almost everywhere. This clearly
demonstrates the need for improvements on the nuclear PDFs. The uncertainties
increase with average transverse momentum $\bar{p}_T$ and towards small values
of $x_\gamma^{\rm obs}$, i.e.\ towards very large values of $x_A^{\rm obs}$,
while they are rather uniformly distributed elsewhere. It is clear that the
EIC measurements (black error bars) represent an improvement of up to an order
of magnitude compared to nCTEQ15-np. The direct contribution
(dot-dashed orange) increases as expected towards large $\bar{p}_T$, in the
backward rapidity region and small $x_A^{\rm obs}$ and is contained in the
highest $x_\gamma^{\rm obs}$-bin. On the other hand, the gluon in the lead ion
(dashed blue line) contributes most at small $\bar{p}_T$ and $x_A^{\rm obs}$,
i.e.\ in the shadowing region, and again in the backward region and at large
$x_\gamma^{\rm obs}$.

\begin{figure}
 \epsfig{file=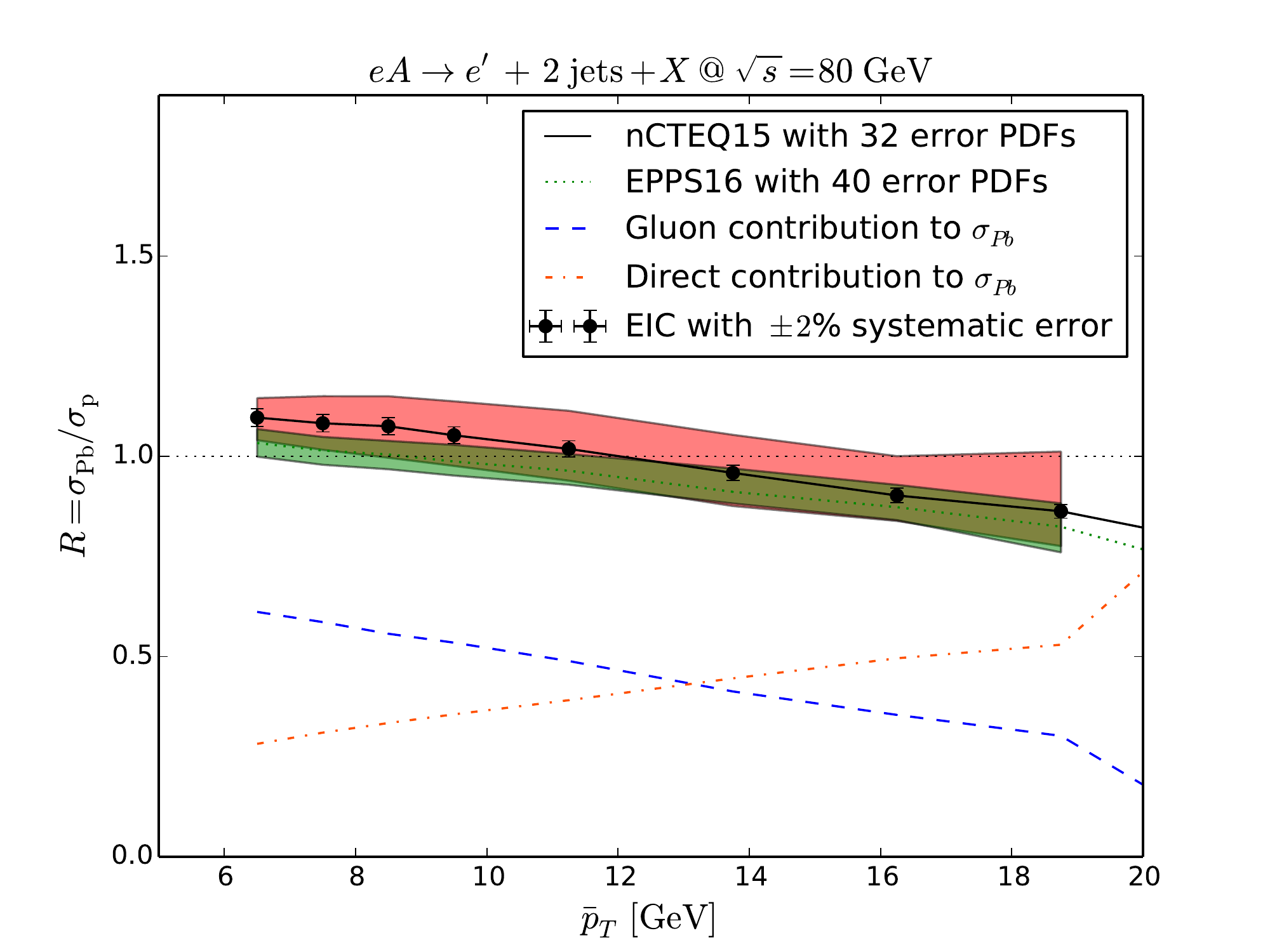,width=0.48\textwidth}
 \epsfig{file=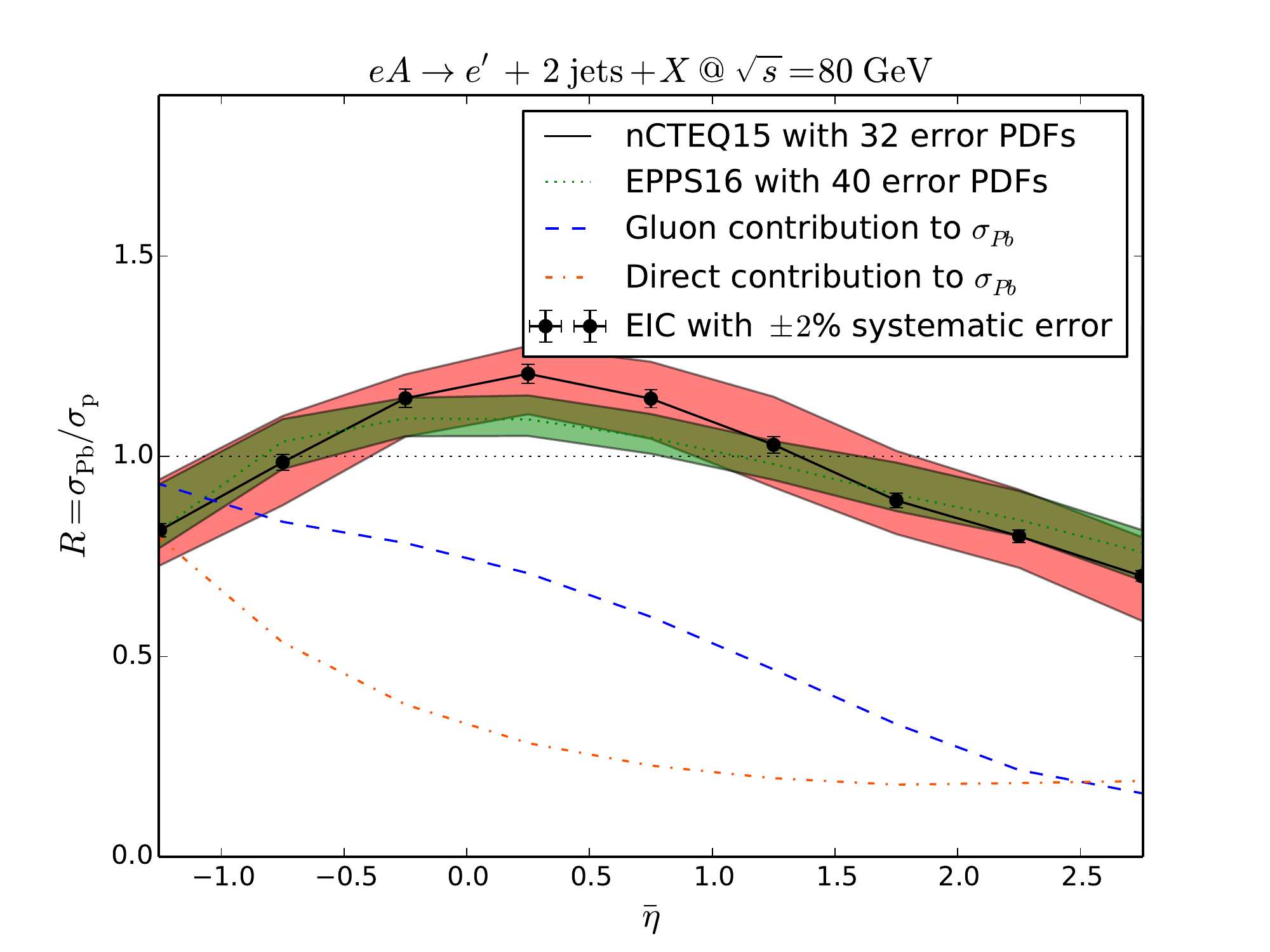,width=0.48\textwidth}
 \epsfig{file=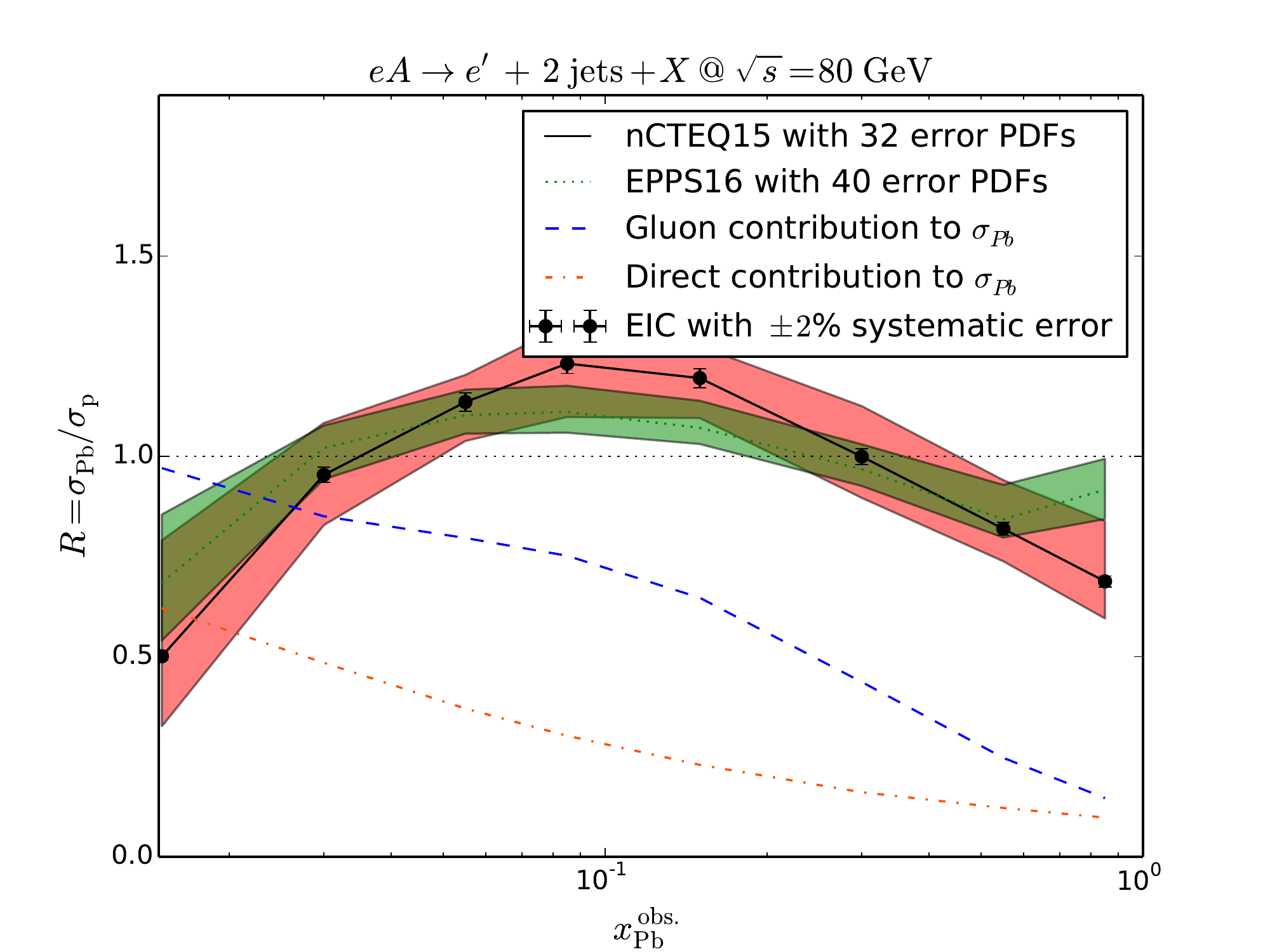,width=0.48\textwidth}
 \epsfig{file=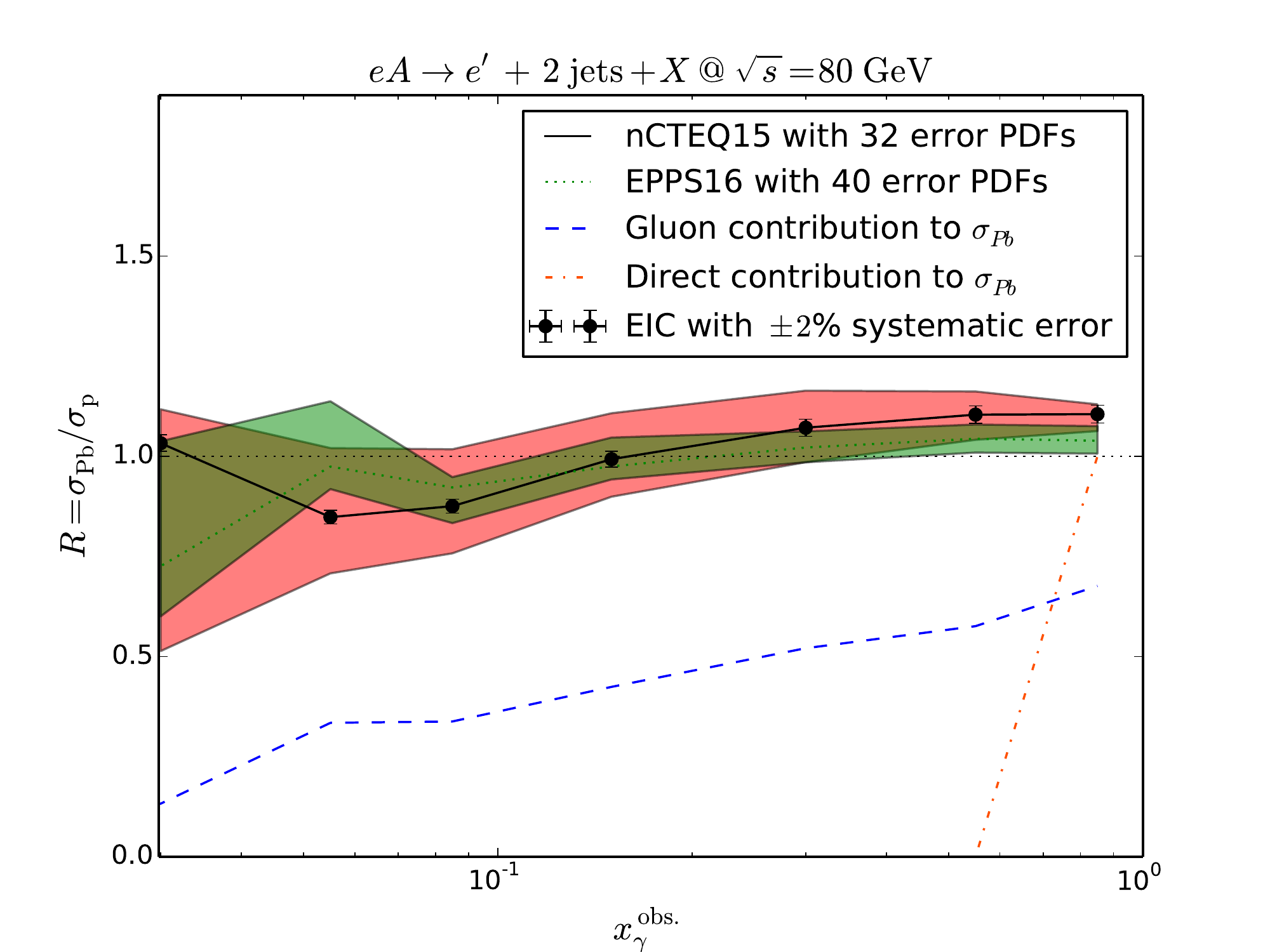,width=0.48\textwidth}
 \caption{\label{fig:4}Same as Fig.\ \ref{fig:3} for the nCTEQ15 fit including
 also inclusive pion data from D-Au collisions at BNL RHIC, and for the central
 EPPS16 fit (dotted green lines) to -- in particular -- dijet data from the
 CERN LHC as well as the corresponding (green-shaded) error bands.}
\end{figure}

When the pion data from BNL RHIC are included, the nCTEQ15 uncertainties are
of course smaller, as it can be seen from Fig.\ \ref{fig:4}. They are then
similar in size to those from EPPS16 (green shaded bands), although one must
keep in mind that these two analyses are based on quite different theoretical
assumptions. It is interesting to see that they nevertheless overlap to a
rather good degree. Even after the inclusion of BNL RHIC pion data
\cite{Adler:2006wg} in nCTEQ15 and CERN LHC, in particular CMS dijet data
\cite{Chatrchyan:2014hqa}, in EPPS16, there is still substantial room for
improvement from the EIC, as the simulated data have error bars that are still
by about a factor of five smaller than the current theoretical uncertainties.


\section{Conclusions and outlook}
\label{sec:5}

In conclusion, after an investigation of inclusive jet production in DIS
\cite{Klasen:2017kwb}, we have studied in this paper the potential of dijet
photoproduction at the EIC to better constrain nuclear PDFs in the near future.
We based our analysis on our theoretical framework of full NLO
\cite{Klasen:1995ab} and approximate NNLO QCD calculations
\cite{Klasen:2013cba}, where we had found little impact of the aNNLO
contributions on the central $K$-factors, but a sizeable reduction of the
scale uncertainty. The latter is also expected to cancel to a large extent in
ratios of ion over free proton cross sections.

Due to the requirements of a minimum transverse momentum of 5, not 4 GeV and
two jets, not only one, the kinematic reach was found to be somewhat smaller
than in inclusive jet DIS. In particular, one cannot expect to reach $x$-values
in the ion down to $2\cdot10^{-4}$ and scales up to $10^3$ GeV$^2$, but only
$x$-values of $10^{-2}$ and $\bar{p}_T^2$ of $400$ GeV$^2$. The jets would be
well contained in a hadronic calorimeter with $\eta\in[-1;3]$

Despite the more limited kinematic reach, we found that one cannot only probe
the EMC and antishadowing regions, but that one can also reach somewhat into
the physically interesting and important shadowing region. Similarly to our
findings in DIS, EIC measurements have the potential to reduce the current
theoretical uncertainty on nuclear PDFs by a factor of 10 to 5, depending on
how much information beyond DIS and DY has been included from existing hadron
colliders.

The implementation of jet mass corrections \cite{deFlorian:2013qia} to our
aNNLO formalism is left for future work. Although they will in particular
introduce a dependence on the jet radius $R$, the impact of these additional
corrections is expected to be even smaller than the one of the aNNLO
contributions as a whole, in particular when $R=1$ as in this study, where
terms $\ln R$ obviously disppear.
Improvements similar to those at the EIC may also be expected from an LHeC
\cite{Paukkunen:2017phq}. Due to its potentially higher center-of-mass energy,
the kinematic reach could even be larger there.
Finally, even transverse-momentum dependent distribution functions (TMDs)
of gluons in protons and nuclei might become accessible in measurements of
dijet asymmetries in polarized or unpolarized $ep$ and $eA$ collisions at the
EIC \cite{Boer:2016fqd}.


\acknowledgments

We thank the organizers of the 8th International Conference on {\it Physics
Opportunities at an ElecTron-Ion-Collider} (POETIC 8), which motivated
this study, for the kind invitation. This work has been supported by the BMBF
under contract 05H15PMCCA. All figures have been produced using
{\tt Matplotlib} \cite{Hunter:2007}.


\bibliographystyle{apsrev}


\end{document}